\documentclass[twocolumn]{aastex631}
\usepackage[toc,page]{appendix}
\usepackage{CJK}
\usepackage{color}

\newcommand{\oiii}{[O~{\sc iii}]} 

\def\simgt{\lower 2pt \hbox{$\, \buildrel {\scriptstyle >}\over {\scriptstyle \sim}\,$}}
\def\simlt{\lower 2pt \hbox{$\, \buildrel {\scriptstyle <}\over {\scriptstyle \sim}\,$}}

\graphicspath{{./}}
\begin{document}
\begin{CJK*}{UTF8}{gbsn}

%\title{Optical Counterparts of the Ultraluminous X-ray Sources in NGC 4631}
\title{Bubble in the Whale: Identifying the Optical Counterparts and Extended Nebula 
for the Ultraluminous X-ray Sources in NGC 4631}

\correspondingauthor{Jianfeng Wu}
\email{wujianfeng@xmu.edu.cn}

\author[0000-0001-9346-3677]{Jing Guo (郭静)}
\affiliation{Department of Astronomy, Xiamen University, Xiamen, Fujian 361005, China}

\author[0000-0001-7349-4695]{Jianfeng Wu}
\affiliation{Department of Astronomy, Xiamen University, Xiamen, Fujian 361005, China}

\author[0000-0001-7584-6236]{Hua Feng}
\affiliation{Department of Astronomy, Tsinghua University, Beijing 100084, China}
\affiliation{Department of Engineering Physics, Tsinghua University, Beijing 100084, China}

\author[0000-0001-8467-6478]{Zheng Cai}
\affiliation{Department of Astronomy, Tsinghua University, Beijing 100084, China}

\author[0000-0002-5683-822X]{Ping Zhou}
\affiliation{School of Astronomy \& Space Science, Nanjing University, 163 Xianlin Avenue, Nanjing 210023, China}
\affiliation{Key Laboratory of Modern Astronomy and Astrophysics, Nanjing University, Ministry of Education, Nanjing 210023, China}

\author[0000-0002-5954-2571]{Changxing Zhou}
\affiliation{Department of Engineering Physics, Tsinghua University, Beijing 100084, China}

\author[0000-0002-0427-9577]{Shiwu Zhang}
\affiliation{Department of Astronomy, Tsinghua University, Beijing 100084, China}

\author[0000-0003-4874-0369]{Junfeng Wang}
\affiliation{Department of Astronomy, Xiamen University, Xiamen, Fujian 361005, China}

\author[0000-0002-0771-2153]{Mouyuan Sun}
\affiliation{Department of Astronomy, Xiamen University, Xiamen, Fujian 361005, China}

\author[0000-0003-3137-1851]{Wei-Min Gu}
\affiliation{Department of Astronomy, Xiamen University, Xiamen, Fujian 361005, China}

\author[0000-0001-7595-1458]{Shan-Shan Weng}
\affiliation{Department of Physics and Institute of Theoretical Physics, Nanjing Normal University, Nanjing 210023, China}

\author[0000-0002-2874-2706]{Jifeng Liu}
\affiliation{Key Laboratory of Optical Astronomy, National Astronomical Observatories, Chinese Academy of Sciences, Beijing 100101, China}
\affiliation{School of Astronomy and Space Sciences, University of Chinese Academy of Sciences, Beijing 100049, China}
\affiliation{WHU-NAOC Joint Center for Astronomy, Wuhan University, Wuhan, Hubei 430072, China}

%\author{et al.}

\begin{abstract}

We present a deep optical imaging campaign on the starburst galaxy NGC 4631 with CFHT/MegaCam. By supplementing the \textit{HST}/ACS and $\textit{Chandra}$/ACIS archival data, we search for the optical counterpart candidates of the five brightest X-ray sources in this galaxy, four of which are identified as ultraluminous X-ray sources (ULXs). The stellar environments of the X-ray sources are analyzed using the extinction-corrected color-magnitude diagrams and the isochrone models. 
We discover a highly asymmetric bubble nebula around X4 which exhibits different morphology in the H$\alpha$ and \oiii\ images. The \oiii/H$\alpha$ ratio map shows that the H$\alpha$-bright bubble may be formed mainly via the shock ionization by the one-sided jet/outflow, while the more compact \oiii\ structure is photoionized by the ULX. We constrain the bubble expansion velocity and interstellar medium density with the MAPPINGS V code, and hence estimate the mechanical power injected to the bubble as $P_w \sim 5\times10^{40}$ erg s$^{-1}$ and the corresponding bubble age of $\sim7\times 10^{5}$ yr. Relativistic jets are needed to provide such level of mechanical power with a mass-loss rate of $\sim10^{-7}\ M_{\odot}\ \rm yr^{-1}$. Besides the accretion, the black hole spin is likely an additional energy source for the super-Eddington jet power. 

\end{abstract}


%%\keywords{Classical Novae (251) --- Ultraviolet astronomy(1736) --- History of astronomy(1868) --- Interdisciplinary astronomy(804)}

\section{Introduction} \label{sec:intro}

Ultraluminous X-ray sources (ULXs) are non-nuclear point-like X-ray sources with isotropic luminosity $L_X\ \gtrsim\ 10^{39}\ \rm erg\ s^{-1}$, which corresponds to the Eddington limit for a $\sim10\ M_{\odot}$ black hole \citep{Feng2011,Kaaret2017}. Two mechanisms are likely to explain the high luminosity: the sub-Eddington accretion onto intermediate-mass black holes (IMBHs) and stellar-mass compact objects undergoing super-Eddington accretion. The minority of ULXs at the higher end of the luminosity range can be explained by the first mechanism, such as ESO~243$-$49 HLX-1 with $L_X\sim 10^{42}~\rm erg~s^{-1}$ \citep{Farrell2009,Webb2012}. Meanwhile, the X-ray spectral properties of most ULXs are consistent with the super-Eddington accretion scenario \citep[e.g.,][]{Gladstone2009,Walton2014,Salvaggio2022}. Recent studies further identified several ULXs powered by neutron stars from the detections of pulsating radiations \citep{Bachetti2014,Furst2016, Israel2017, Israel2017b, Weng2017,Carpano2018, Wilson-Hodge2018, Sathyaprakash2019, Rodrguez2020ApJ, Quintin2021}.

The definitive approach to decipher the nature of non-pulsating ULXs is the dynamical mass measurement of the accretors, which relies on the optical spectroscopy of the donor stars. However, the archived optical data on ULXs are far less abundant than X-ray data because most of the ULX optical counterparts are very faint ($m_V > 21\ \rm mag$) and located in fairly crowded regions. Previous studies found that most of the ULXs are associated with young star clusters, showing the donor stars might be the OB type \citep{Roberts2008, Poutanen2013}. For a limited number of ULXs, the nature of the donor stars are unambiguously identified (e.g., M101 ULX-1 and NGC 7793 P13), while the dynamical studies on these systems supported the stellar-mass accretor scenario \citep{Liu2013,Motch2014}. 

A number of ULXs have surrounding bubble nebulae detected from deep optical imaging observations \citep[e.g.,][]{Pakull2002, Ramsey2006, Soria2010, Soria2021}, the majority of which are considered to be formed via shock ionizations driven by the interactions of strong jets/outflows and the ambient interstellar medium (ISM). Strong outflows may be ubiquitous for ULXs under supercritical accretion \citep[e.g.,][]{Narayan2017,Weng2018,Zhou2019,Qiu2021,Kosec2021}. The kinetic power and age of the bubble can be inferred from its size and expanding velocity \citep{Weaver1977}, and hence may reveal the kinematics of jets/outflows and the accretion physics of ULXs \citep{Pakull2010,Cseh2012,Soria2021}. For the other few cases, the high-ionization features (e.g., He~{\sc ii}~$\lambda4686$) in the spectra of the optical nebulae imply that the photoionization could be the major origin of the extended structure \citep{Pakull2002}. Both shock ionization and photoionization may certainly be working at the same time while dominating different parts of the same optical nebula \citep{Gurpide2022,Zhou2022}. 

In this work, we report on an optical broad-band and narrow-band imaging campaign for the Whale Galaxy NGC 4631 to identify the optical counterparts and surrounding extended nebulae of the ULXs, for which \citet{Soria2009} presented a detailed study of their X-ray properties. As a late-type starburst galaxy 7.35~Mpc away, NGC 4631 (Figure~\ref{fig:rgb}) has been extensively studied in multiwavelengths. The existence of molecular outflows, abundant gas and the X-ray halo reveals the diversity 
of objects and astrophysical processes \citep[e.g.,][]{Yama2009,Irwin2011,Melen2015}. From the archival \textit{XMM-Newton} data, \cite{Soria2009} identified five brightest X-ray sources scattered in NGC 4631 and found that four of them (X1, X2, X4, X5) can be classified as ULXs.\footnote{While \citet{Mineo2012} classified X4 as a high mass X-ray binaries in the sub-Eddington state based on the \textit{Chandra}-measured luminosity, we adopt \citet{Soria2009}'s classification throughout this work.} For the purpose of studying their physical nature and stellar environments, we analyze the optical images of all five X-ray sources in this paper combining the Canada-France-Hawaii Telescope (CFHT) and \textit{Hubble Space Telescope} (\textit{HST}) observations, supplemented with \textit{Chandra} data to determine the precise astrometry. The details of the five X-ray sources can be found in Table~\ref{tab:coord}.

This paper is organized as follows.
In Section~\ref{sec:obser} we present the optical and X-ray observations and data reduction. In Section~\ref{sec:optcount}, we improve the relative astrometry and identify optical counterpart candidates for the X-ray sources, which are investigated in Section~\ref{sec:CMD} based on their locations on the isochrone diagrams. In Section~\ref{sec:bubble}, we present a newly discovered bubble nebula around X4 and the analyses on its morphology and kinetic power. Section~\ref{sec:conclusion} summarizes our conclusions.

%X2 is aligned with a young star cluster with high extinction.

\begin{deluxetable*}{cccccccc}
\tablenum{1}
\tablecaption{List of the Five Brightest X-ray Sources in NGC 4631\label{tab:coord}}
\tablewidth{0pt}
\tablehead{
\colhead{Source ID} & \colhead{R.A.} & \colhead{Dec.} & \colhead{\textit{Chandra} Net Counts}  &
\colhead{Off-axis} &  \colhead{Opt-X Error Circle}
&  \colhead{$N_{\rm H}$}   &   \colhead{E(F606W-F814W)}\\
\colhead{         } & \colhead{(J2000)} & \colhead{(J2000)} & \colhead{(0.5--8.0 keV)}  &
\colhead{(arcmin)} & \colhead{(arcsecond)} & \colhead{($10^{21}$~cm$^{-2}$)} & 
}
%\decimalcolnumbers
\startdata
X1 & 12 42 15.99 & +32 32 49.47 & 6.7$\pm{2.8}$ & 4.10 & 0.673   &2.4$^{+0.3}_{-0.3}$    & 0.35$^{+0.15}_{-0.15}$\\
X2 & 12 42 11.12 & +32 32 35.63 & 981.2$\pm{32.9}$ & 3.05 & 0.275   &28.3$^{+3.6}_{-3.2}$   & 3.80$^{+1.74}_{-1.55}$\\
X3 & 12 42 06.13 & +32 32 46.43 & 357.6$\pm{19.6}$ & 2.11 & 0.269   &2.0$^{+1.0}_{-0.9}$    & 0.29$^{+0.48}_{-0.43}$\\
X4 & 12 41 57.42 & +32 32 02.79 & 77.7$\pm{9.2}$   & 0.19 & 0.280   &0.32$^{+1.02}_{-0.32}$    & 0.05$^{+0.49}_{-0.16}$\\
X5 & 12 41 55.57 & +32 32 16.77 & 2977.8$\pm{55.8}$ & 0.51 & 0.268   &2.0$^{+0.2}_{-0.2}$   & 0.29$^{+0.10}_{-0.10}$\\
\enddata
\tablecomments{The $N_{\rm H}$ values are retrieved from \citet{Soria2009} which were obtained from the \textit{Chandra} spectral analyses.}
\end{deluxetable*}

\begin{deluxetable*}{cccccccccccl}
\tablenum{2}
\tablecaption{Observation Log of NGC 4631\label{tab:obslog}}
\tablewidth{0pt}
\tablehead{
\colhead{Instrument} & &\colhead{Source ID} & &\colhead{ObsID} & &\colhead{Filter} &  &\colhead{Observation Date (UT)} & & \colhead{Exposure Time}
}
%\decimalcolnumbers
\startdata
CFHT/MegaCam & & X1-X5 & & 20AS01 & & H$\alpha$   & & 2020-03-23 & & 12$\times$900 sec \\
{(PI: Jing Guo)          } & &  {   } & &        & & \oiii        & & 2020-05-19 & & 5$\times$750 sec \\
{          } & & {   } & &        & & $u$           & & 2020-03-23 & & 5$\times$126 sec \\
{          } & & {   } & &        & & $g$           & & 2020-03-23 & & 5$\times$126 sec\\
{          } & & {   } & &        & & $r$           & & 2020-03-23 & & 5$\times$126 sec\\
\hline
\textit{HST}/ACS      & & X1,X2,X3 & & j8r331010 & & F606W & & 2003-08-03  & &  676 sec \\
{          } & & X1,X2,X3 & & j8r331020 & & F814W & & 2003-08-03  & &  700 sec \\
{          } & & X4, X5 & & j8r332010 & & F606W & & 2004-06-09  & &   676 sec \\
{          } & & X4, X5 & & j8r332020 & & F814W & & 2004-06-09  & &   700 sec \\
\hline
\textit{Chandra}/ACIS & & X1-X5   & & 797  & &  & &  2000-04-16  & & 60 ksec \\
\enddata
%\tablecomments{ \textit{Chandra} data}
\end{deluxetable*}

\begin{deluxetable*}{cccccc}
\tablenum{3}
\tablecaption{List of the reference stars\label{tab:refstar}}
\tablewidth{0pt}
\tablehead{
\colhead{Reference ID} & \colhead{X-ray Coordinates} & \colhead{Optical Coordinates} & \colhead{Off-axis}  &\colhead{ Net Counts}&
\colhead{X-ray positional error}  \\
\colhead{  } & \colhead{\textit{Chandra}} & \colhead{\textit{HST}} &  \colhead{(arcmin)}&\colhead{\textit{Chandra}}  &
\colhead{(arcsecond)}
}
%\decimalcolnumbers
\startdata
Ref.1 & 12 42 25.78 +32 33 21.40 & 12 42 25.79 +32 33 21.23&6.2 & $120\pm 11$ & 0.32 \\
Ref.2 & 12 42 04.03 +32 34 08.60 &12 42 04.03 +32 34 08.41 & 2.7 & $107\pm 10$ & 0.19 \\
\enddata
\tablecomments{The \textit{HST} coordinates are given by the Dolphot package.}
\end{deluxetable*}

\section{Observations and Data reduction} \label{sec:obser}

\subsection{CFHT} \label{subsec:cfht}
We obtained optical broad-band and narrow-band imaging of NGC 4631 with the 3.6-m CFHT located 
on Mauna Kea, Hawaii. The MegaCam instrument mounted on CFHT has a wide field of view (1~deg$^2$) which can fully cover all the 5 luminous X-ray sources in NGC 4631. The detector consists of 40 CCDs, each of which has $2048\times4612$ pixels with $0\farcs187\times 0\farcs187$ per pixel. 
We are awarded a total of 4.5-hour exposure time (PI: Jing Guo, ObsId: 20AS01) executed in 2020 March and June. The images are taken with three broad bands 
($u$, $g$, and $r$) and two narrow bands (H$\alpha$ and \oiii). The H$\alpha$ and \oiii\ filters have a width of $\sim100$~\AA, centered at 6590~\AA\ and 5006~\AA, respectively. A dithering pattern was applied during the observations to cover the CCD gaps, which requires at least five exposures for each band. 
The detailed observation log is listed in Table \ref{tab:obslog}.

\begin{figure*}
\epsscale{1.15}
\plotone{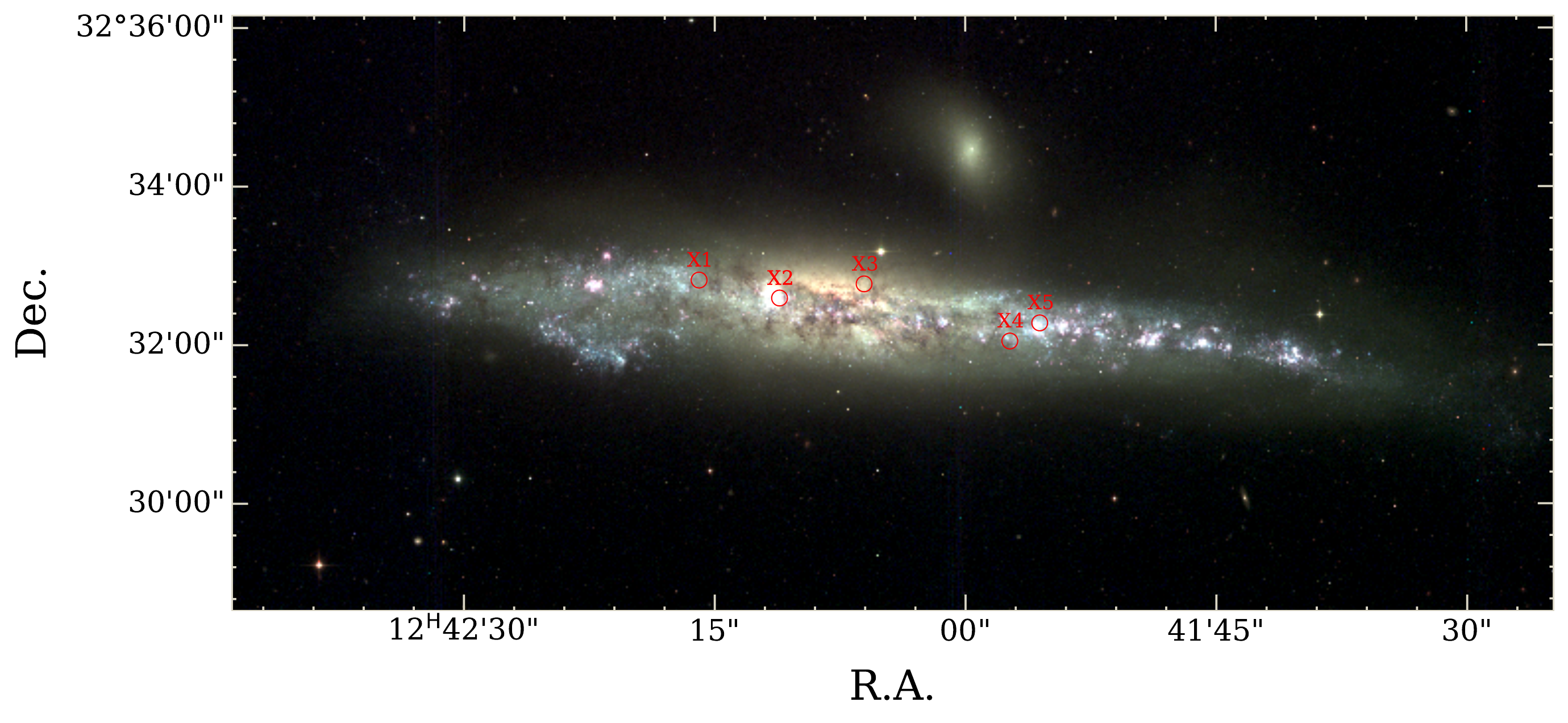}
\caption{The RGB-like image combines five CFHT/MegaCam filters, including the three broad bands ($u$, $g$, $r$) and two narrow bands (H$\alpha$, \oiii). The $u$, $g$, and $r$ bands are shown in blue, green, and red colors, respectively, while the H$\alpha$ and \oiii\ filters are represented by crimson and teal colors, respectively. The red circles label the positions of the five X-ray sources. }\label{fig:rgb}
\end{figure*}

The data products we received have been preprocessed with the \verb+Elixir+ pipeline, which includes bias-subtraction, flat-fielding, etc., for each individual frame \citep{Magnier2004}. The first step is to perform precise astrometric calibration and to stack images from individual exposures in the same band for the purposes of eliminating  CCD gaps and reaching the desired sensitivity level. In the stacking procedure, \verb+SExtractor+ \citep{Bertin1996} was applied on each image of single exposures to generate the catalog of all
point sources with coordinates. The astrometric solutions were then computed with the \verb+SCAMP+ \citep{Bertin2006} software by referencing the catalog 
from {\it Gaia} Data Release 1 (DR1). We utilized the \verb+SWarp+ \citep{Bertin2010} task to perform the image stacking.
The astrometric error during these processes are $< 0\farcs03$.
A multi-color image of NGC 4631 is shown in Figure~\ref{fig:rgb}. This RGB-like image combines the three broad bands and two narrow bands. The five red circles
label the positions of the five luminous X-ray sources analyzed in \citet{Soria2009}. X3 is more likely a black hole X-ray binary in its high/soft state, while the remaining four X-ray sources are classified as ULXs, among which X1 is a supersoft ULX. 
%Now we have got the complete image without gaps in each band and can start to do photometry.

%%For the edge-on character of NGC 4631, it is diffcult to find the point optical counterpart of ULXs in CFHT image. The zoom in images of 5 ULXs are shown in fig. 2. 

\begin{figure*}
\epsscale{1.1}
\plotone{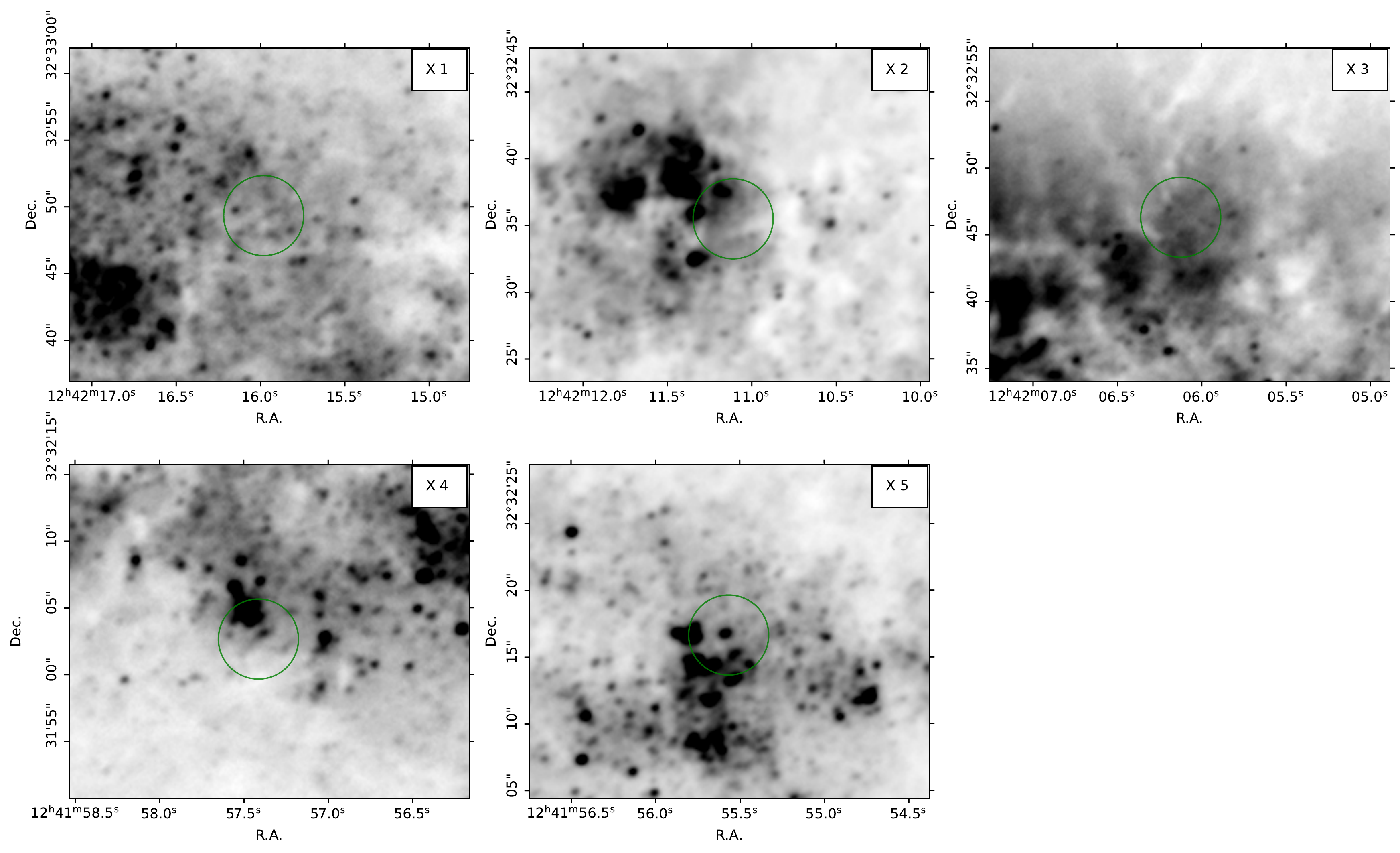}
\caption{The CFHT/MegaCam $g$-band images of X1-X5 and their vicinity, respectively. The green circles are centered at the X-ray location of each source with a radius of $3\arcsec$. Optical counterparts are difficult to identify in these broad-band images due to the seeing limit  ($0\farcs45$--$0\farcs75$)  for the ground-based CFHT.}\label{fig:cfht}
\end{figure*}

We select 40 point sources from the Pan-STARRS1 DR2 catalog \citep{Flewelling2018,Flewelling2020} that are isolated and have an appropriate magnitude (17--19 mag) to serve as photometry references. The Pan-STARRS1 DR2 catalog does not flag the source whether it is a star. Thus we select such a relatively large set of referencing sources aiming to obtain a more statistically reliable photometry calibration. We adopt the conversion equations from Pan-STARRS filters to MegaCam filters provided by the Canadian Astronomy Data Centre (CADC),\footnote{https://www.cadc-ccda.hia-iha.nrc-cnrc.gc.ca/en/megapipe/docs/filt.html} except for H$\alpha$ we use the formula provided by \citet{Boselli2018}. Finally, for each given source, we can derive an array of magnitude values calibrated from the 40 reference stars. The peak value of the best-fit Gaussian profile to the histogram of the magnitude values was adopted as the measured magnitude for this source.

\begin{figure*}
\epsscale{1.1}
\plotone{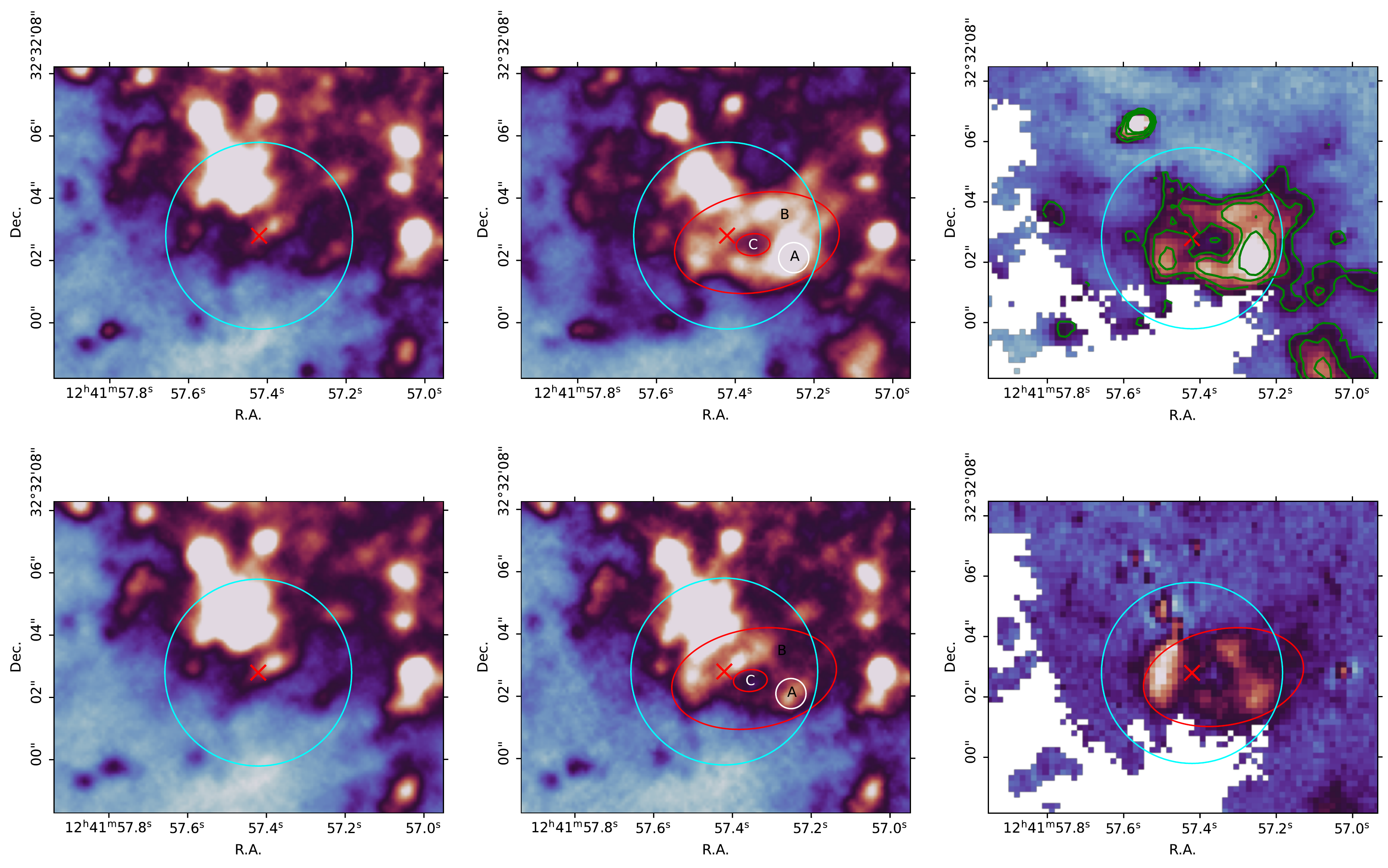}
\caption{From left to right in the first row, the first panel is the CFHT/MegaCam $r$-band image, where the cyan circle is centered at the X-ray location (the red cross symbol) of X4 with a $3\arcsec$ radius. The half length of the red cross represents the X-ray positional error of X4. The second is the H$\alpha$ image, residing with a bubble-like structure around X4. The brightest region is marked as A region in the white circle. When performing the photometry for the whole bubble, the shape is adopted as the region between the two red ellipses, marked as B (which includes the A region). The cavity in the center is marked as C. The third panel is the result of subtracting the underlying continuum component from the H$\alpha$ image. Most of the stellar sources have been removed here. The blank regions represent the dropped pixels that do not have adequate $S/N$. In the second row, the images of $g$, \oiii\ and \oiii\ with continuum removed are shown in turn. }\label{fig:x4bubble}
\end{figure*}

The zoom-in CFHT/MegaCam $g$-band images of the five luminous X-ray sources are shown in Figure~\ref{fig:cfht}. For the stacked image in each band, we estimate the exposure depth reaching 25--26 mag~arcsec$^{-2}$ at $3\sigma$ level. Due to the seeing limit of ground-based imaging, it is difficult to identify the exact optical counterparts to the X-ray sources at their crowded locations in the edge-on galaxy NGC 4631. However, in the H$\alpha$ narrow-band images we discover a bubble-like extended nebula surrounding X4, which may be inflated by the jet or wind launched from the ULX accretion disk (Figure~\ref{fig:x4bubble}). The projected size of this bubble structure is $\sim130$~pc $\times$ 100~pc. 
%, i.e., the net photon counts of a source (N$_{\rm s}$) must be above the galaxy background (N$_{\rm bkg}$) with at least 3 times of $\sqrt{\rm N_{\rm bkg}}$

%However, this extended source is embedded in the galaxy background. 
To obtain a more precise profile of the extended bubble structure, we need to subtract the continuum contribution from the H$\alpha$ image. \citet{Boselli2018} 
utilized a large set of unsaturated stars and derived an empirical equation (see their Eqn. 4) to relate the $g-r$ color and the H$\alpha$ magnitude. Following \citet{Boselli2018}, we use the data products which were processed by CADC with \verb+MegaPipe+ upon our request. 
\verb+MegaPipe+ will subtract the sky background, normalize the flux in the whole stacked image, and provide a catalog of detected point sources \citep{Gwyn2008}. We filtered out the pixels with low signal-to-noise ratio  ($S/N \leqslant 5$), and then applied the equation in \citet{Boselli2018} pixel by pixel. The generated image is shown in the upper right panel of Figure~\ref{fig:x4bubble}. Most of the point sources around X4 have been removed from the H$\alpha$ image. The morphology of the extended structure are clearly revealed. 

For the \oiii\ narrow-band image, we derived a similar equation connecting the \textit{g-r} color and the \oiii\ magnitude by roughly assuming the continuum magnitude is linearly related to wavelength in the given range (i.e., the continuum follows a power-law spectral profile; see details in Appendix~\ref{Oiii continuum}):
\begin{equation}\label{eqn:oiii}
     m_{g/\rm [O~{\sc III}]}\ \approx\  m_g - 0.155\times (m_g-m_r), 
\end{equation}
where $m_{g/\rm [O~{\sc III}]}$ is the magnitude of the continuum that falls within the \oiii\ narrow-band filter. After applying the equation pixel by pixel, we obtain an \oiii\ narrow-band image for which most of the continuum contribution has been eliminated (see the lower right panel of Figure \ref{fig:x4bubble}). As for the continuum-subtracted H$\alpha$ image, the point sources have been mostly removed from the \oiii\ image, proving the efficacy of our continuum subtraction method. We will discuss this bubble structure in details in Section~\ref{sec:bubble}. 

%%In order to get a clear extent structure in H$\alpha$ band. We need subtract the continuum from the H$\alpha$ image. We requested CADC to help us restack those images, in which the background of galaxy is removed. And from a empirical method\citep{Boselli2018}, the point source can be removed in the H$\alpha$ image  The subtracted images for each source can be found in figure . As you can see, most point sources have been erased. And the science disscussion is in Sec.\ref{sec:ident}.

%%A multicolor image is shown in Figure \ref{fig:rgb}. This RGB like image is combined with 3 broad band and 2 narrow band. And the positions of 5 ULXs is pointed out by the red circles.

%To get the point optical counterparts, we reexamine the archived Hubble data.

\subsection{HST} \label{subsec:hst}
\begin{figure*}
\epsscale{0.9}
\plotone{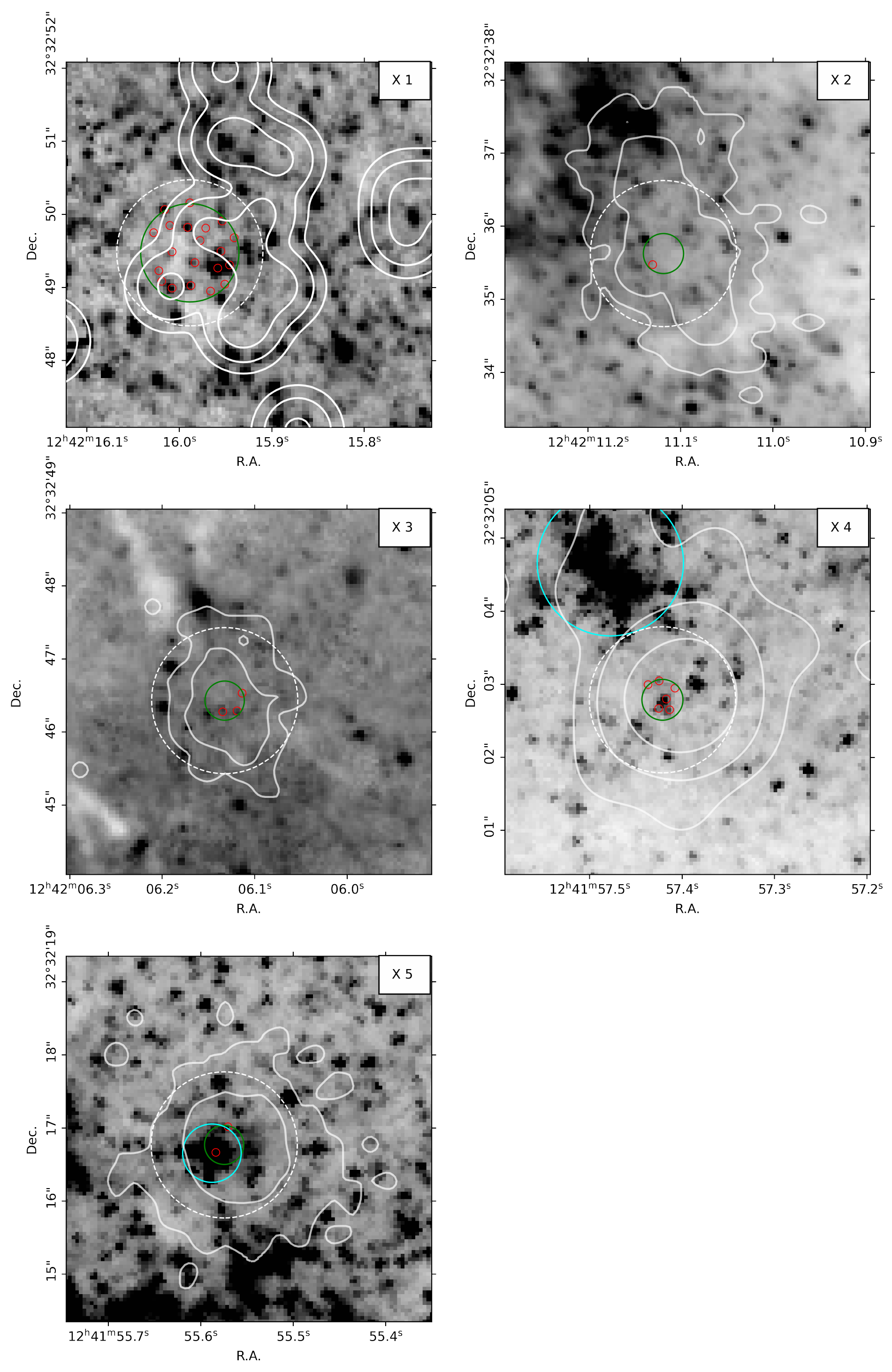}
%\plotone{fig4.png}
\caption{The {\it HST}/ACS F606W images of each X-ray source, with the overlaid white contours representing the X-ray flux level from the {\it Chandra}/ACIS data (contours not in uniform scales among the five panels). In each panel, the green circle is centered at the X-ray location with the radius represents the respective error circle. The numbered red circles are the optical counterpart candidates of the X-ray source. The white dashed circle has a radius of $1\arcsec$. The cyan circle in the middle right panel marks the young star associations northeast to X4, while that in the bottom left panel labels the compact young star group associated with X5. }\label{fig:hst_chandra}
\end{figure*}

NGC 4631 has been observed with the Advanced Camera for Surveys (ACS; \citealt{Ford1998}) onboard {\it HST} (see Table~\ref{tab:obslog}). 
%To study the evolution of NGC 4631, this galaxy was observed with HST/ACS several times to inspect its disk and halo. 
The five luminous X-ray sources were completely covered by the observations in Proposal 9765.
The field containing X1, X2, and X3 was observed in 2003 August (ObsID j8r331010 for the F606W filter and 
j8r331020 for F814W), while X4 and X5 were covered by the observations in 2004 June (ObsID j8r332010 for 
F606W and j8r332020 for F814W). Each observation has a total exposure time of 1376 sec. 
The images of the five X-ray sources in the F606W band are shown in Figure~\ref{fig:hst_chandra}. 
%Residing in the edge of the observation field, the image around X1 looks smaller than other sources images.

We aim to identify the optical 
counterparts of X-ray sources with the \textit{HST} imaging and derive their magnitudes. Astrometric calibration is also needed for the \textit{HST} images. As the lack of coverage upon the galaxy disk of NGC 4631 in \textit{Gaia}, we are not able to directly align \textit{HST} images with the \textit{Gaia} references. CFHT images with a large field of view are reused as the reference images to align the \textit{HST} data. We selected seven reference sources in each  \textit{HST} observation to perform astrometric calibration, for which we obtained the RMS residual of $0\farcs03$. Then we employed the \verb+Dolphot+ package to perform Point Spread Function (PSF) photometry. \verb+Dolphot+ can identify point sources 
in heavily crowded areas and return their Vega magnitudes \citep{Dolphin2000}. The \textit{acsmask} task was used to
flag bad pixels, and the \textit{calcsky} task can calculate the sky background. After these preprocessing, the PSF photometry was accomplished by the \textit{dolphot} task. The parameters in \textit{dolphot} are configured referring to \citet{Williams2014} where they made a series of artificial stars to test a mesh grid parameters and found out the most suitable parameter set for crowded fields. 

%The point sources located in the error circle are considered as optical counterparts candidates of ULXs. Error circle can be derived from the combination of X-ray and optical data as described in sec. \ref{sec:optcount}. We show the photometry result for the candidates. The result is discussed in section \ref{sec:optcount}
%{\bf We also use Sextractor, Scamp to complete the astrometry.(?)} 

\subsection{\textit{Chandra}} \label{subsec:chandra}

In the X-ray band, NGC 4631 has been observed by \textit{Einstein}, \textit{ROSAT}, \textit{Chandra}, and \textit{XMM-Newton}. To obtain precise locations of the X-ray sources, we reprocessed the \textit{Chandra}/ACIS data which have a sub-arcsec angular resolution. The \textit{Chandra} observation (ObsID 797) was carried out on 2000 April 16 for a total of 60 ksec exposure time. We perform X-ray astrometry and photometry in this work. The spectral and timing properties of these X-ray sources were presented in details in \citet{Soria2009}. 

The data were reprocessed with the CIAO (v4.13) package. The \verb+chandra_repro+ task 
was applied to create a new $level=2$ event file calling the latest calibration products (CALDB v4.9.4) and 
more advanced algorithms. For astrometric calibration, we aligned the \textit{Chandra}/ACIS images to the \textit{HST}/ACS images (see Section~\ref{sec:optcount} for details). 
%After obtaining an initial X-ray point source list using \verb+wavdetect+, we aligned the \textit{Chandra}/ACIS-S image with \textit{HST}/ACS image to correct for absolute %astrometry. Five common sources existing in both images were selected based on whether they are isolated 
%and bright enough. Then the \verb+wcs_match+ task was used to projecting the X-ray 
%coordinate system onto the \textit{HST}/ACS image. The RMS astrometric residual after the transformation is $\approx0.3$~arcsec. 

The CIAO script \verb+deflare+ was used to remove the background flares ($>3\sigma$) which only accounts for $\approx4\%$ of the total exposure time. 
The full-band (0.5--8.0~keV) X-ray image was then generated using the \textit{ASCA} grade 0,2,3,4,6 events. The PSF and exposure maps were produced accordingly. The final X-ray point source detection was carried out using \verb+wavdetect+. The detection threshold is set to be $10^{-6}$, while the wavelet scales are 1, $\sqrt{2}$, 2, $2\sqrt{2}$, and 4 pixels. The coordinates return by \verb+wavdetect+ are adopted as the X-ray positions for the five luminous X-ray sources.

\section{Identifying the optical counterparts} \label{sec:optcount}

To identify the optical counterparts of the X-ray sources, we improve the astrometry of \textit{Chandra}/ACIS images relative to the \textit{HST}/ACS images following the methodology laid out in \citet{Yang2011}. Because of the small field of view of \textit{HST}/ACS, only one common source can be registered from the \textit{Chandra} to the \textit{HST} images. Therefore, we supplement the \textit{HST}/ACS observation on an adjacent and partly overlapping field (ObsID jc9l04010) to taking a mosaic image using the \verb+AstroDrizzle+ package. The second common source is therefore added. These two objects are identified as point X-ray sources \citep{Wang2016, Evans2010}, and their coordinates and other information are listed in Table \ref{tab:refstar}. We use the CIAO task \verb+wcs_match+ to register the \textit{Chandra} image to the \textit{HST} image. The RMS residual is $0\farcs02$. The updated positions of five X-ray sources are listed in Table~\ref{tab:coord}. 
% The first object is identified as a point X-ray source in \citet{Wang2016}. It has also been detected in the optical \citep{Eckart2005} and near-infrared \citep{Cutri2014} bands. \verb+Dolphot+ gives its \textit{HST}/ACS coordinates as R.A. = 12h42m04.03s, Dec. = +32d34m08.4s (J2000). It has $107\pm10$ photons detected by CIAO/\verb+wavdetect+ which provides an error radius of $\sim0\farcs1$ at 90\% confidence. The X-ray position of the reference object in \textit{Chandra}/ACIS image is at R.A. =  12h42m04.03s, Dec. = +32d34m08.7s (J2000). 

We calculated the 95\% \textit{Chandra} positional error radius for each source using Equation 5 in \citet{Hong2005} which has considered the PSF variations across the field of view. We then converted it to the 1$\sigma$ error radius by applying the relation $r_X = r_{X(95\%)}/1.95996$ in \citet{Zhao2005}. The size of the error circle is primarily related to the number of counts and the off-axis angle of the source.  %The relative rotation between the optical and X-ray images is another source of the positional error, which cannot be corrected with only one reference object. We estimate the systematic rotation accuracy between \textit{Chandra}/ACIS and Gaia DR1 in a few \textit{Chandra}/ACIS observations which cover multiple active galactic nuclei. The average rotation value is $\sim2\arcmin$. Therefore, the error introduced by the rotation can be calculated from the distance between the X-ray sources and the reference object. 
Finally, we adopt the positional uncertainty of each source as the quadratical combination of all kinds of errors: the average X-ray positional error of the two reference objects ($0\farcs19$), the X-ray positional error of each X-ray source ($0\farcs15$-$0\farcs63$), the error caused by the alignment between the \textit{HST} and \textit{Chandra} images, and the error of optical coordinates which were generated during the astrometric calibration of \textit{HST} and CFHT images. The latter two kinds of errors are both ignorable compared to the X-ray positional errors. The final positional uncertainties of the five X-ray sources are listed in Table~\ref{tab:coord}.

In Figure~\ref{fig:hst_chandra}, we overlay the X-ray flux contours (solid white lines) onto the \textit{HST}/ACS/F606W images for each X-ray source. The green circles represent the uncertainties of X-ray positions. X1 has the largest error circle ($0\farcs66$) because of the small number of \textit{Chandra} net counts ($\approx7$). There are multiple candidate optical counterparts for X1 detected by \verb+Dolphot+, which are labeled by small red circles in the upper left panel of Figure~\ref{fig:hst_chandra}. The error radii of X2--X5 are similar ($ \sim 0\farcs3$). X2 has one candidate optical counterpart in its X-ray positional error circle, while both X3 and X4 have a few candidates in their respective error circles. X5 is located within a crowded region while two individual sources are detected in the error circle. We will analyze these candidate optical counterparts and the surrounding stellar environments in the next section.

\begin{figure*}
\epsscale{0.95}
\plotone{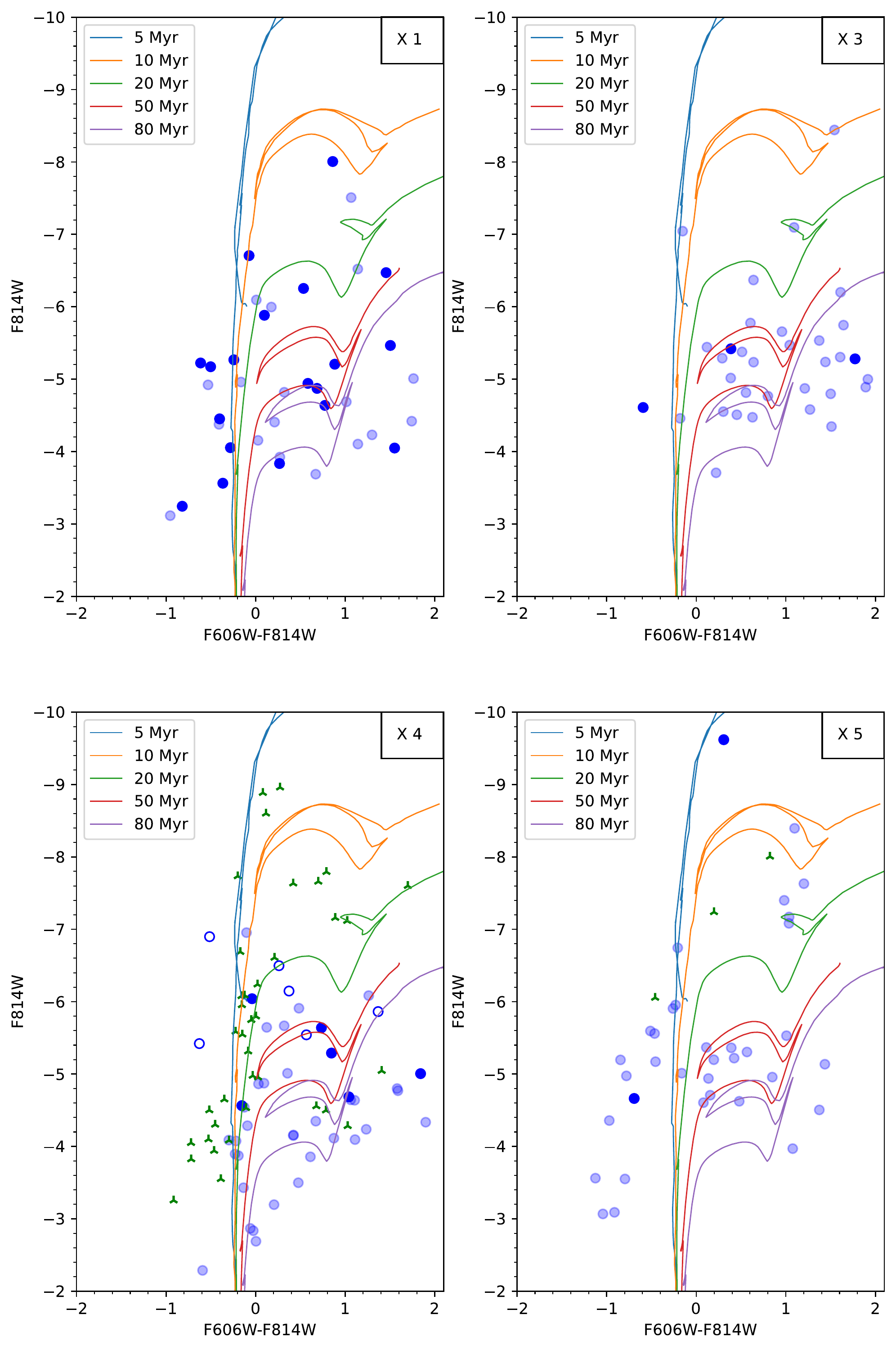}
\caption{The color-magnitude diagrams (CMDs) for optical point sources around X1, X3, X4, and X5, respectively. The solid blue dots are the optical counterpart candidates within the error circle. The light blue dots are the point sources within $1\arcsec$ but outside the error circle which could be born in the same environment. The green $tri\_up$ symbols in left-lower panel represent the sources in the star group northeast of X4. Extinction correction has been applied based on the X-ray hydrogen column density. The open blue circles labels the loci of the optical counterpart candidates of X4 when the extinction value is adopted from the \textit{XMM-Newton} spectral modeling.}\label{fig:iso}
\end{figure*}

\section{Color-Magnitude Diagram} \label{sec:CMD}

For most ULXs, the optical emission is dominated by the X-ray reprocessing on the accretion disk \citep{Tao2011}. Nevertheless, the optical Color-Magnitude Diagrams (CMD) can be used to infer the age of the stellar environments around ULXs, which could potentially suggest the nature of the ULX donor stars. The Padova Stellar Evolution Code (PARSEC; \citealt{Bressan2012}) provides the isochrone databases for almost all the mainstream telescope filters.\footnote{http://stev.oapd.inaf.it/cgi-bin/cmd}  Here we utilize the isochrones based on the \textit{HST}/ACS filters system.

We derive the extinction $A_V$ from the hydrogen column density obtained by \citet{Soria2009} via X-ray spectral analyses of the \textit{Chandra} observations based on the relation of 
$N_{\rm H}$ (cm$^{-2}$) $= (2.21\pm0.09) \times 10^{21}$ $A_V$ (mag) presented in \citet{Guver2009}. To convert $A_V$ to the extinction in the \textit{HST}/ACS filter system $E{\rm (F606W-F814W)}$, we then interpolate the central wavelengths of these filters to the extinction law derived in \citet{Cardelli1989}. The extinction value for each X-ray source is listed in Table~\ref{tab:coord}.

Closely aligned with a young stellar cluster, X2 has large extinction $E{\rm (F606W-F814W)}=3.8$~mag, which may introduce significant uncertainties when applying the CMD to derive the ages of its surrounding stars. Therefore, we only plot the isochrones for the other four X-ray sources (Figure~\ref{fig:iso}). For each panel, the solid blue dots stand for the optical counterpart candidates of the X-ray source.  The light blue dots represent the stars within $1\arcsec$ (36 pc; see the white dashed circles in Figure~\ref{fig:hst_chandra}) from the X-ray position but outside the optical-to-X-ray error circle. 

The immediate surrounding stars of X1 do not appear to be closely associated like in a star group or cluster. Its optical counterpart candidates, as well as the nearby stars within $1\arcsec$, span a wide range of age from 5 Myr to 80 Myr, which indicates that they are unlikely born at the same time or in the same environment. As discussed in Section~\ref{sec:optcount}, the positional error circle of X1 is also much larger ($0\farcs673$), corresponding to $\approx25$~pc. 
For X2, the sole optical counterpart shown in Fig~\ref{fig:hst_chandra} is likely to be not reliable because of the large extinction.
For X3, the three optical counterpart candidates have ages of $\sim$50--80 Myr which are consistent with the ages of the environmental sources.
% appears like a very young OB-type star ($<5$~Myr) in the CMD, while most of the surrounding stars have older ages of $\sim50$--80 Myr. This indicates the dominance of disk emission in the optical band for X3. 
For X4, the ages of the surrounding stars range mostly in $\sim20$--80 Myr. The six candidate optical counterparts also show similar ages. There appears to be a star association 
northeast of X4 with the size of $\approx2\arcsec$ across (71~pc). The CMD shows that most of its member stars are very young with ages of 5--20 Myr (the green $tri\_up$ symbols in the lower left panel of Figure~\ref{fig:iso}). It is worth noting that the $N_{\rm H}$ value of X4 derived from the \textit{XMM-Newton} spectral modeling %(3.6$^{+0.6}_{-0.6}\times10^{21}$ cm$^{-2}$)
is one order of magnitude higher than that with the \textit{Chandra} data \citep{Soria2009}. The optical counterpart candidates of X4 would be younger, $<$ 20 Myr (see fig~\ref{fig:iso}), if the {XMM-Newton} extinction value were adopted. X5 appears to locate within a compact star group ($\approx0\farcs8$ across; corresponding to 28~pc). Two point sources are identified within the error circle with ages of $\sim$ 5 Myr, while three more individual sources in this star group are resolved by \verb+Dolphot+, which have ages $\sim5$--10~Myr (the green $tri\_up$ symbols in Figure~\ref{fig:iso} lower right panel). Therefore, X5 is likely associated with a young star cluster. 

It is worth noting that the extinction derived from the hydrogen column density obtained with X-ray spectra represent an upper limit for the candidate optical counterparts and their surrounding stars. If significant intrinsic absorption exists for the X-ray source, the extinction would be much smaller. A conservative lower limit would be the Galactic extinction along the light of sight of NGC 4631, which is $A_V = 0.015$~mag \citep{Schlafly2011}, corresponding to $N_{\rm H} = 3.3\times 10^{19}$~cm$^{-2}$. This is ignorable compared to that from the X-ray spectroscopy. The true values of extinction should be in between the above lower and upper limits. Overestimation of the extinction would place the stars at younger age regions on the CMD. However, it is probably reasonable to assume that the candidate optical counterparts and surrounding stars would suffer from high extinction (i.e., close to the upper limits), since NGC~4631 appears as an edge-on disk galaxy.

\begin{figure}
\epsscale{1.35}
\plotone{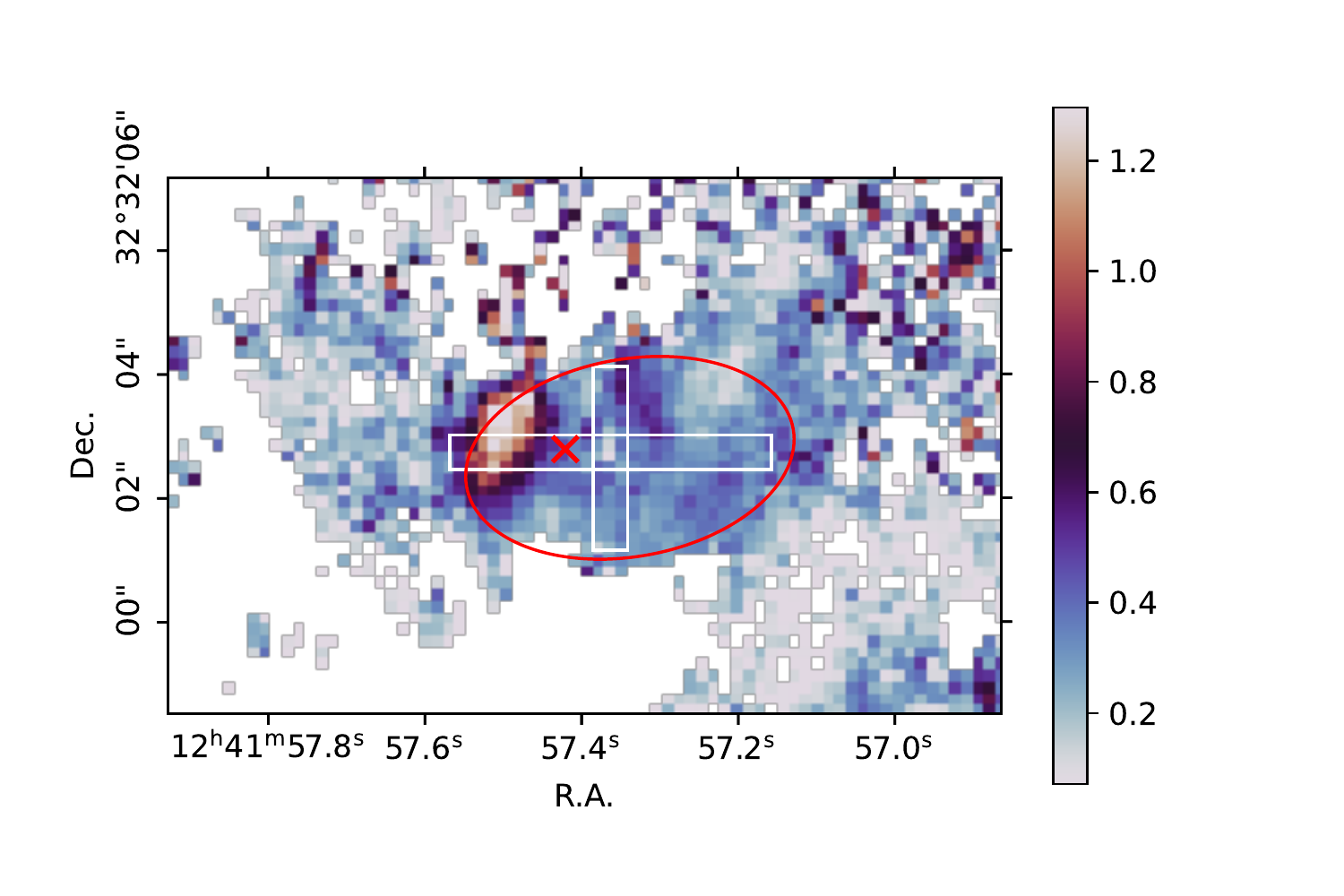}
\caption{The \oiii/H$\alpha$ flux ratio map for the extended structure around X4, where the vertical color bar shows the line ratio values. The red ellipse represents the profile of the H$\alpha$ bubble, while the red cross marks the position of X4. The white horizontal and vertical bars illustrate the major and minor axes of the extended structure.}\label{fig:ratOiiiHa}
\end{figure}

\section{A Newly Discovered Bubble Structure Around X4}\label{sec:bubble}

\subsection{Morphology Analysis}\label{sec:morph}

Both of the continuum-subtracted H$\alpha$ and \oiii\ images display a clear extended structure around X4 (see the right two panels in Figure~\ref{fig:x4bubble}), while exhibiting different morphology in the two bands. In the H$\alpha$ image, the structure appears more like an inflated bubble with the size of $\sim130$~pc $\times$ 100~pc. The X-ray source X4 is not located in the center. Instead, this H$\alpha$ bubble structure appears to be sourced from the location of the ULX (see the red cross in Figure~\ref{fig:x4bubble}) and is oriented toward the southwest direction, reaching maximum luminosity in the outermost region, $\sim100$ pc away from X4. The extended nebula in the \oiii\ image has a smaller size. In contrast to the H$\alpha$ bubble, the brightest region of the \oiii\ structure is to the east of X4, and is substantially closer to the X-ray source ($\simlt25$ pc). 

\begin{figure}
\epsscale{1.3}
\plotone{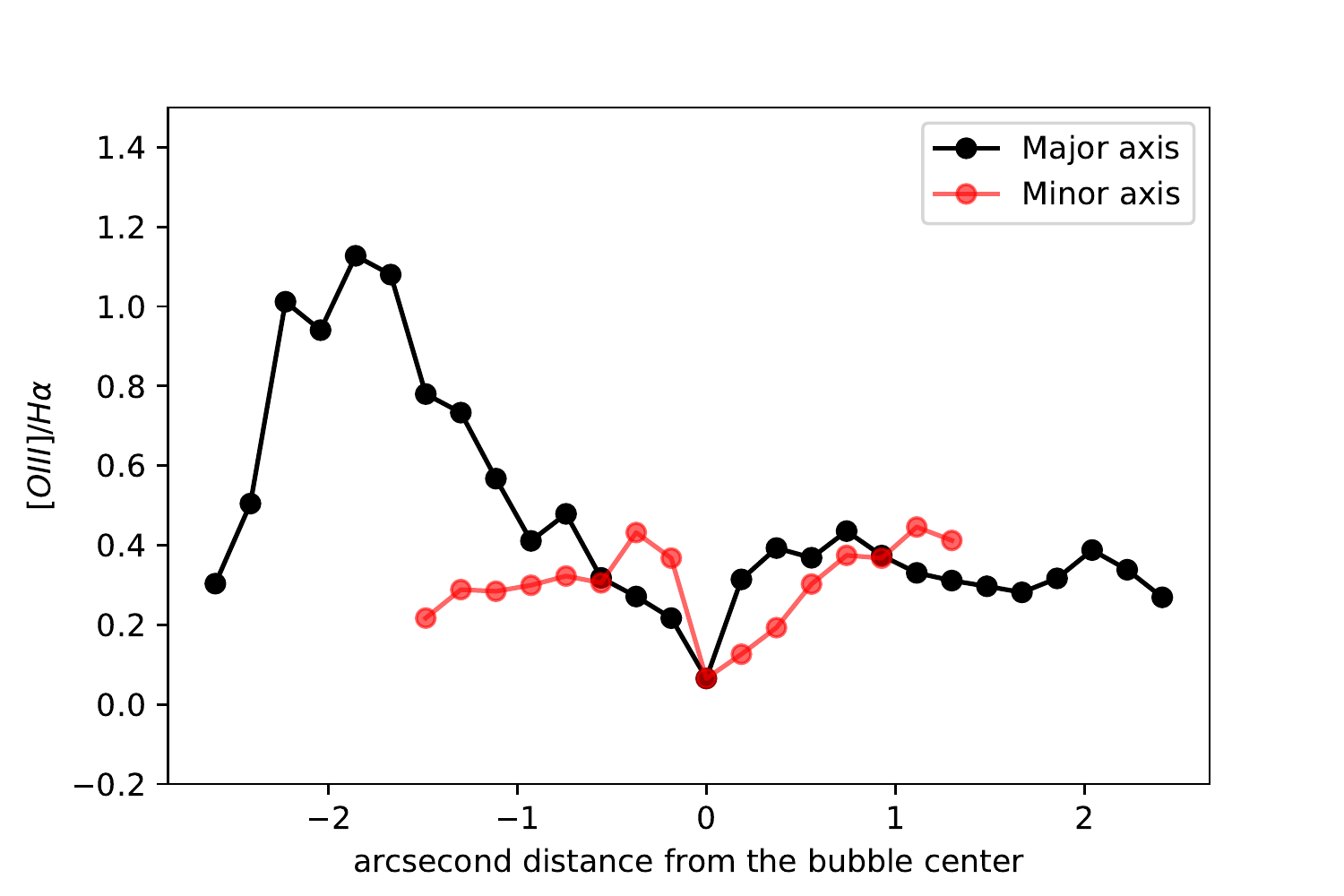}
\caption{The spatial profile of the \oiii/H$\alpha$ flux ratio along the major and minor axes of the extended nebula (see the white bars in Figure~\ref{fig:ratOiiiHa}). }\label{fig:Oiii_Ha_two}
\end{figure}

The extended structures around ULXs may originate from photoionization or shock ionization, both of which could coexist while playing major roles in different parts of the structure \citep{Moon2011,Gurpide2022,Zhou2022}. Generally, in the photoionization process, the line flux ratio \oiii/H$\beta$ tends to peak at or near the ionizing source and declines outwards. For the shock-ionized bubble, the edge region has higher excitation level and exhibits the higher \oiii/H$\beta$ ratio than in the central area. Here we use the \oiii/H$\alpha$ ratio as a proxy, since it is reasonable to assume that the line ratio $H\alpha/H\beta\equiv \tau$ remains constant in the bubble area. The typical $\tau$ value is $\sim3$ for ULX bubbles \citep{Allen2008}. We will calculate the exact value for our case in the next subsection. Derived from the continuum-subtracted \oiii\ and H$\alpha$ images, the \oiii/H$\alpha$ ratio map is shown in Figure~\ref{fig:ratOiiiHa}, in which the red ellipse marks the bubble shape in the H$\alpha$ band (same as that in the upper middle panel of Figure~\ref{fig:x4bubble}). We extract the \oiii/H$\alpha$ line ratio roughly along the major and minor axes of the bubble (the white horizontal and vertical bars in Figure~\ref{fig:ratOiiiHa} respectively) and obtain a clearer spatial profile, which is illustrated in Figure~\ref{fig:Oiii_Ha_two}. The \oiii/H$\alpha$ ratio reaches its minimum in the bubble center and increases outwards along both axes, which suggests that the H$\alpha$ bubble is mostly dominated by the shock ionization. There is a bump of the \oiii/H$\alpha$ ratio in the east edge of major axis, coinciding with the brightest \oiii\ region. This area is likely formed predominantly by photoionization. It is indeed close to the ULX which is presumably the source of ionizing photons. The peak  \oiii/H$\alpha$ value in this area is $\simgt 1$. Combined with the calculated $H\alpha/H\beta$ line ratio of $\tau\sim3.75$ (see Section~\ref{sec:mecha}), the peak \oiii/H$\beta$ would be $\sim4$, which is similar to that of photoionization-dominated nebulae found in previous ULX bubble studies (e.g., \citealt{Soria2021}). 

The shock-ionized bubbles around ULXs can be formed via two mechanisms: through explosive events like supernovae (i.e., supernova remnants) or being inflated by continuous jet/outflow from ULXs \citep{Pakull2005}. However, ULX bubbles often have sizes of a few hundred pc \citep[e.g.,][]{Ramsey2006,Grise2011}, which are one order of magnitude larger than normal supernova remnants. Although there exists the possibility of very energetic hypernova explosions, it is unlikely considering the stellar environments and the survival of ULXs as binary systems during the events \citep{Feng2011}. Therefore, we suggest the H$\alpha$ bubble structure around X4 is more likely to be inflated by the ULX jet/outflow.  

It is worth noting that, unlike many other ULX bubbles, this extended structure around X4 only has a one-sided lobe to the southwest direction, while X4 itself is close to the east edge of the bubble. The missing of the lobe in the other direction is not caused by the relativistic beaming because the bubble expansion velocity $v_s$ is only at the order of hundred $\rm km\ s^{-1}$, far below the speed of light. For example, the bubble around the ULX in NGC~5585 has an expansion velocity of $125\ \rm km\ s^{-1}$ \citep{Soria2021}. For this bubble around X4, the expansion velocity is estimated to be $v_s\sim110$ $\rm km\ s^{-1}$ (see Section~\ref{sec:mecha}). 
%Therefore, it is likely that the ISM to the northeast is too tenuous to interact with the jet/wind, resulting in no bubble forming.

The asymmetric profile of an extended nebula could imply the density gradient of ISM or outflows, as suggested for IC~342 X-1 \citep{Cseh2012}. Here in our case, one viable scenario is that the ISM is much denser to the east of X4, resulting in an outflow blocked by the dense medium. Hence, the east area is mainly photoionized by the ULX itself (as shown by the \oiii/H$\beta$ ratio profile), while most other regions are dominated by shock ionization through the outflow. Similar situations can be found in the simulations of supernova feedback \citep{Creasey2011, Pardi2017}. In their simulations, if the ISM density reaches $10^2$--$10^4$ cm$^{-3}$ and the ejection temperature is lower than 10$^6$ K, the injected energy of the outflow will be immediately lost due to the strong radiative cooling in the high density regions, dubbed the overcooling problem. 
%However, even with a young star cluster located to the northeast of X4 which may enhance the density of ISM, the density gradient is unlikely to rapidly decline to the typical value ($\sim1$ cm$^{-3}$) within $\sim$ 100 pc.

An alternative interpretation of this unusual morphology is that the accretion disk of X4 has launched an asymmetric outflow, i.e., the outflow to the east direction is much weaker or nonexistent. There have been numerical simulations showing that an asymmetric or even one-sided outflow can be formed from the accretion disk if the accretor is rotating and is accompanied with a complex magnetic field \citep{Lovelace2010, Dyda2015}.

It is also possible that both of the two mechanisms are responsible for this asymmetric morphology. The side with the weaker/absent outflow has more dense ambient ISM, leading to photoionization dominating the compact east area close to the ULX, while the shock-ionized bubble is only formed to the opposite direction.
%In all the possible interpretations, both photoionization and shock ionization play important roles in the bubble formation.

%(move to the conclusion section as future perspective) We do not find any radio counterpart at or surrounding X4 in the archival Very Large Array observations in 1988 January, suggesting the jet may be discontinuous. Optical spectroscopic observations are needed to determine the expanding velocity of the bubble and can help us derive the line ratio such as [Si {\sc ii}]/H${\alpha}$, which can clarify the shock ionization and photo ionization.

\subsection{Mechanical Power Estimation}\label{sec:mecha}

To estimate the mechanical power needed to inflate the bubble, we first calculate the H$\alpha$ luminosity from the surface brightness of the structure measured with \verb+Python/Photutils+. The brightest region in the H$\alpha$ band (marked with ``A" in Figure~\ref{fig:x4bubble}) has a surface brightness of $19.34\pm0.01$ mag arcsec$^{-2}$. The whole bubble structure, which is confined in a donut shape (region B in Figure~\ref{fig:x4bubble}, subtracting region C while including region A), has an average surface brightness of 19.64 $\pm$ 0.01 mag arcsec$^{-2}$. Using the surface brightness and bubble size $R$, We can estimate the injected mechanical power $P_w$. 

Based on the standard bubble theory \citep{Weaver1977,Pakull2005}, $P_w$ can be calculated as 
\begin{equation}\label{eqn:power}
P_{39} \approx 3.8{R_2}^2{v_2}^3n \ \rm{erg\ s^{-1}} ,
\end{equation}
where $P_{39} \equiv P_w/(10^{39}\ \rm erg\ s^{-1}) $; $R_2 \equiv R/(100\ \rm pc)$; $v_2 \equiv v_s /(100\ \rm km\ s^{-1})$; $n$ is the ISM number density in unit of cm$^{-3}$. As the ULX is located at the edge of the $H\alpha$ bubble, we conceive that the one-sided jet/wind formed this one-lobe bubble. Therefore, we adopt the scale of the whole bubble to substitute the radius, i.e., $R = 130\ \rm pc$ and $R_2 = 1.3$. The number density $n$ can be derived from the H$\beta$ luminosity $L_{H\beta}$, the expanding velocity $v_s$, and the area of the spherical bubble $A$ \citep{Dopita1996}, 
\begin{equation}\label{eqn:density}
    n = 1.3\times10^5L_{H\beta}A^{-1}v_2^{-2.41}\ \rm cm^{-3}. 
\end{equation}
%we assume a velocity range of $\sim100$--300 km s$^{-1}$ based on the model of \citet{Siwek2017} and previous observations on other bubbles around ULXs.

\begin{figure}
\epsscale{1.2}
\plotone{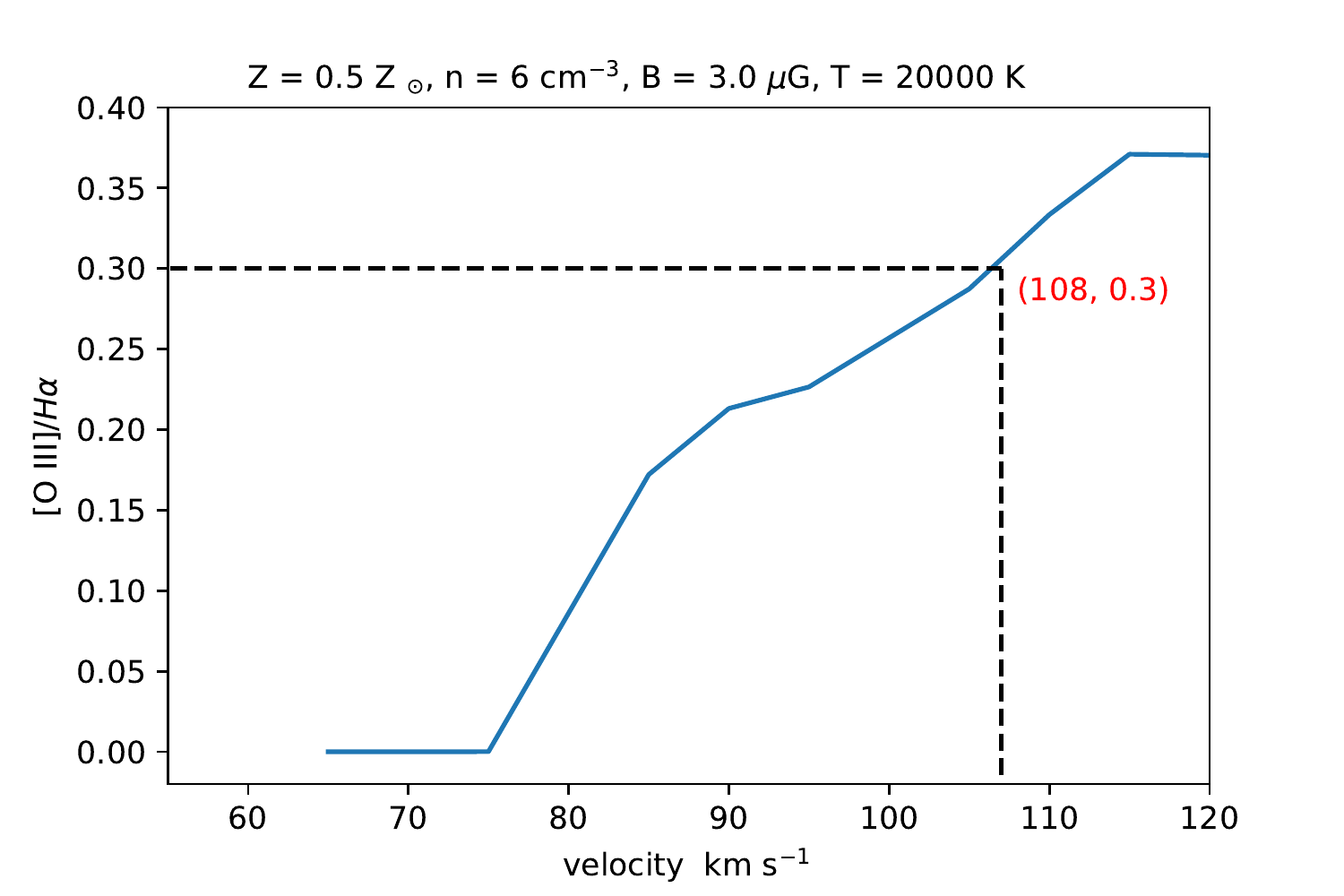}
\caption{The adopted solution from MAPPING V code. The ISM number density is $6\ \rm cm^{-3}$. When the shock velocity $v_s\ \approx$ 108 $\rm{km\ s^{-1}}$, the \oiii/H$\alpha$ flux ratio is consistent with the observed value of $\approx0.3$. }
\label{fig:mapping}
\end{figure}

The surface area $A$ of the spherical bubble with a diameter of 130 pc is calculated as $5\times10^{41}$ cm$^2$. Without available H$\beta$ imaging, the H$\beta$ luminosity can be derived from H$\alpha$ luminosity using the Balmer line ratio $\tau$. The intrinsic H$\alpha$ luminosity is $L_{H\alpha}=1.2\times10^{38}$ erg~s$^{-1}$, calculated from the bubble surface brightness. With the lack of optical spectroscopy on the bubble, the precise expanding velocity $v_s$ is not available. We employed the widely used shock-ionization model MAPPINGS V \citep{Allen2008} to estimate $\tau$ and $v_s$. We carried out a series of calculations with MAPPINGS V which returned values for a variety of line ratios to compare with the observation. We fixed the metallicity to 0.5 solar abundance \citep{Pilyugin2014}, magnetic field to a typical value of 0.3 $\mu$G, and arranged a large input grid of the shock velocity $v_s$ and hydrogen number density $n$, finding a set of solution matching the observed \oiii/H$\alpha$ line ratio, which is $\approx0.3$ along the most parts of the H$\alpha$ bubble (see Figure~\ref{fig:Oiii_Ha_two}). The result is shown in Figure~\ref{fig:mapping}. We obtained the following parameter values in this solution: $n\sim6\ {\rm cm^{-3}}, v_s\sim110\ \rm{km\ s^{-1}}$, and the Balmer line ratio $\tau\sim3.75$.
%$\alpha/\rm H\beta\ \sim 3.75$ }

Combining Eqns.~(\ref{eqn:power}) and (\ref{eqn:density}), we can derive a relation where the mechanical power $P_w$ is determined by the H$\alpha$ luminosity $L_{H\alpha}$, the line ratio $\tau$ and expanding velocity $v_s$:
\begin{equation}\label{eqn:p39}
    P_{39} \approx 5.0\times10^5R_2^2(L_{H\alpha}/\tau) A^{-1}v_2^{0.59}.
\end{equation}
By substituting the MAPPING V solution, we calculated the value of $P_{39} \approx51$, i.e., the injected mechanical power $P_w \sim 5\times 10^{40}$ erg~s$^{-1}$. The lifetime $t$ of the bubble is estimated to be $t= \frac{3}{5}R/v_s\sim7\times 10^5$ yr. 
%The result is illustrated in the left panel of Figure~\ref{fig:P39}. The green shaded area marks the range of possible $\tau$ and $v_2$, while the colorbar represents the  $P_{39}$ value. This plot shows that the injected mechanical power $P_w$ is likely in the range of $(7.5$--$12.5)\times 10^{39}$ erg~s$^{-1}$. 

\begin{figure}
\epsscale{1.2}
\plotone{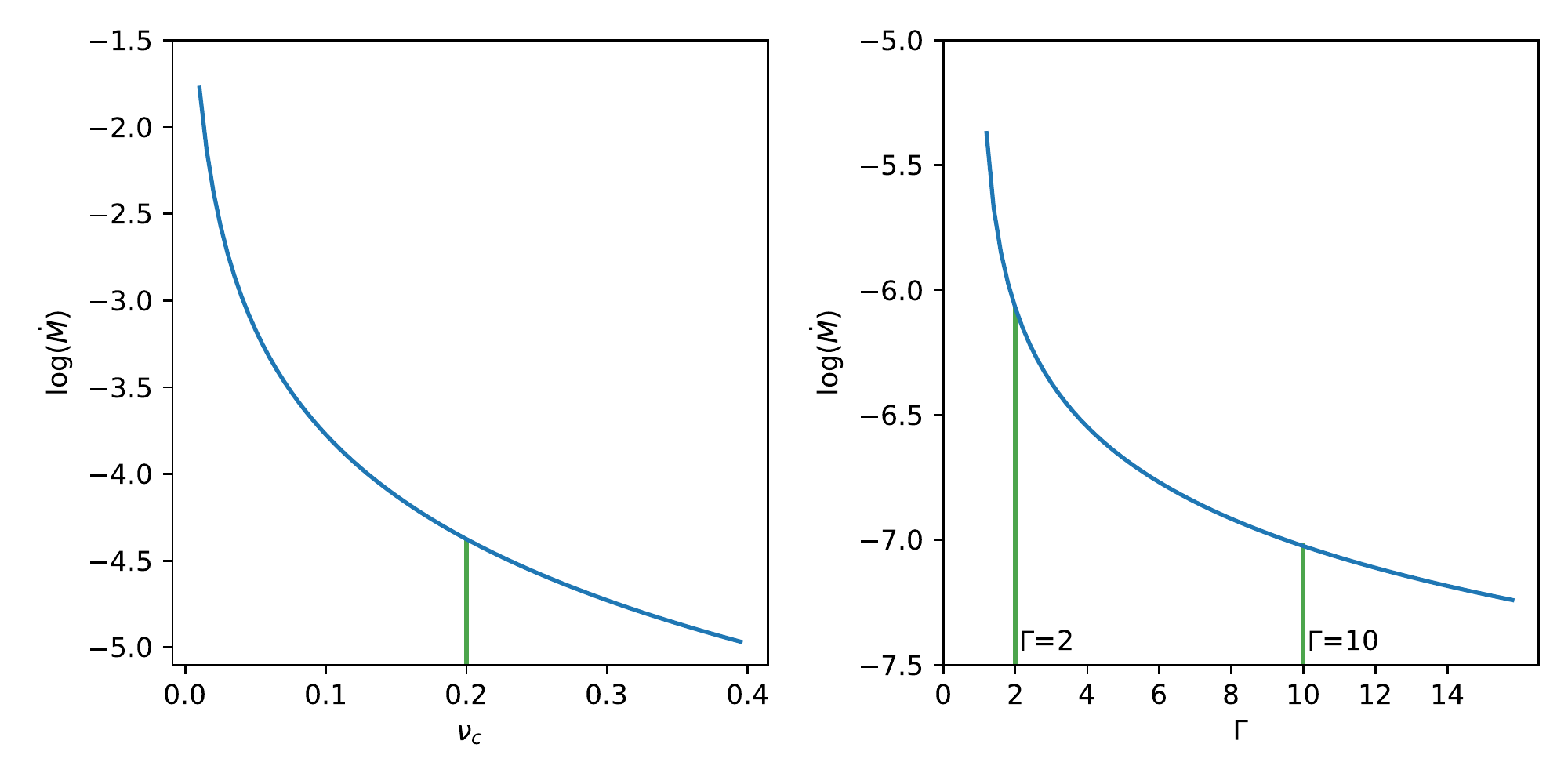}
%The left panel represents the mechanical power $P_w$ injected to the bubble. The $x$-axis is the bubble expansion velocity, and the $y$-axis is the ratio $\tau = L_{H\alpha}/L_{H\beta}$. The shaded green area marks the possible parameters space. The colorbar illustrates the scale of the mechanical power, while the white dashed lines labels the $P_{39}$ value. 
\caption{The estimated mass-loss rate $\dot{M}$ of non- and mildly-relativistic wind (left panel) and the relativistic jet (right panel). The $x$-axis is the wind velocity normalized by the speed of light $v_c$ in the left panel and the bulk Lorentz factor $\Gamma$ in the right panel.}
\label{fig:P39}
\end{figure}

We can now calculate the mass-loss rate $\dot{M}$ of the ULX jet/wind. In the non- and mildly-relativistic scenario, the injected power can also be expressed as $P_w = \frac{1}{2}\dot{M}v_w^2$ \citep{Weaver1977}, where $v_w$ is the velocity of jet/wind. This equation can be transformed to 
\begin{equation}\label{eqn:massrate}
    \dot{M}\approx3.3\times10^{-8}P_{39}/v^2_c\ \ M_{\odot}\ {\rm yr}^{-1},
\end{equation}
where $v_c (\equiv v_w/c)$ is the wind velocity in unit of the speed of light $c$. The left panel of Figure~\ref{fig:P39} shows the range of the mass-loss rate $\dot{M}$ and its dependence on wind velocity $v_c$. With a typical value of $v_c\sim0.2$ \citep{Pinto2016,Pinto2021,Kosec2018}, the mass-loss rate is calculated as $\sim 10^{-5}  M_{\odot}\ {\rm yr}^{-1}$ (the green vertical line in the left panel of Figure~\ref{fig:P39}), which means that $\sim 10\ M_{\odot}$ will be lost through ULX wind in the bubble lifetime. Such a high mass loss rate will make it difficult to sustain the long-term stable accretion activity. 

Instead, we consider the jet-powered bubble scenario with highly relativistic ejection velocity. The mass-loss rate $\dot{M}$ can be inferred with the following equation \citep{Kaiser1997,Cseh2012},
\begin{equation}
    \dot{M} = \frac{P_w}{(\Gamma-1)c^2}.
\end{equation}
Adopting a minimum bulk Lorentz factor $\Gamma = 2$, we can derive the $\dot{M}$ ranges as $\sim 10^{-6}\ M_{\odot}\ \rm yr^{-1}$, while for a higher bulk Lorentz factor, like $\Gamma = 10$, the mass-loss rate would decrease to $\sim 10^{-7}\ M_{\odot}\ \rm yr^{-1}$  (Figure~\ref{fig:P39} right panel), which is clearly more realistic for sustaining the accretion of ULXs. This would suggest that relativistic jets are necessary to generate the shock-ionized ULX bubbles like the one we found around X4. Mildly-relativistic winds with typical velocity of $\sim0.2c$ alone would not provide adequate mechanical power. Steady jets at the distance of NGC 4631 would have a radio flux level of $\sim1\ \mu$Jy, which is difficult to detect with current facilities, while flaring jets that are 1--2 orders of magnitude brighter could be detected with, e.g., the Karl G. Jansky Very Large Array (VLA), like the case of Holmberg~II X-1 \citep{Cseh2015}. We have searched the VLA database; sensitive radio imaging data with sub-arcsec resolution on NGC 4631 will be publicly available in the near future.

The estimated jet mechanical power of NGC 4631 X4 is greater than its radiative luminosity. It would also be above the Eddington limit if the accretor mass is less than $\sim100\ M_\odot$, as for most ULXs. This is similar to the cases of several microquasars found in nearby galaxies, e.g., NGC~7793 S26 \citep{Pakull2010} and M83 MQ1 \citep{Soria2014}, both of which have the same level of jet power at $\sim10^{40}$~erg~s$^{-1}$. The Galactic microquasar SS~433 also has the jet power far exceeding its X-ray luminosity \citep{Fabrika2004}. These microquasars also have surrounding shock-ionized bubble structures detected with optical/infrared emission lines. Their X-ray luminosity are admittedly orders of magnitude below the canonical definition of ULXs. However, they could have had episodes of super-Eddington radiative luminosity in the past, while NGC 4631 X4 itself also had sub-Eddington X-ray luminosity ($\sim10^{37}$~erg~s$^{-1}$) during its \textit{Chandra} observation. Furthermore, the low X-ray luminosity of SS~433 is also due to the heavy obscuration along the line of sight; only reflected X-ray flux is detectable (e.g., \citealt{Begelman2006,Middleton2021}). On a much larger scale, some powerful Fanaroff-Riley~II radio galaxies and blazars have been found to have jet power much greater than the radiative luminosity \citep{Ito2008,Ghisellini2014}. NGC 4631 X4 and the aforementioned microquasars appears to be analogs of these active galaxies at stellar scales.

%The estimation of mechanical power is based on the assumption that the bubble is formed by the one-sided jet. However, the asymmetry of bubble can also be created by double-sided jet injecting to interstellar medium with a magnetic field. Under this circumstances, the bubble radii $R_2\sim 0.65$, the mechanical power $P_{w}$ is calculated as $12\times 10^{39}$ erg~s$^{-1}$. Consequently, the mass-loss} rate $\dot{M}$ in the non-relativistic scenario would be $1\times 10^{-5}\ M_{\odot}\ {\rm yr}^{-1}$. While when considering the relativistic effect, the mass-loss} rate is calculated as $\sim 10^{-7}\ M_{\odot}\ {\rm yr}^{-1}$ in a minimum Lorentz factor $2$. That a rate is also justifiable to support the accretion in ULXs. And the reconsidered lifetime of bubble $\sim 3\times 10^5$ yr, which imply the bubble is one of the youngest in the ULXs nebula group. Under this scheme, the bubble is younger, brighter than other bubble, and is likely be formed by the relativistic jet.}

From another perspective, we consider the energy sources of the injected mechanical power. In case of all the jet mechanical power originates from the accretion, i.e., the release of gravitational potential energy of the accreted material, the needed accretion rate $\dot{m}$ can be calculated from $P_w = \epsilon \dot{m}c^2$., where $\epsilon$ is the fraction of accretion power converted into mechanical energy. Under the assumption of $\epsilon=0.1$, which is already considered as exceptionally high, the needed accretion rate is $\dot{m}\sim 10^{-5}\ M_{\odot}\ \rm yr^{-1}$. For more realistic $\epsilon$ values, the needed accretion rate would be even higher. This would suggest that there should be additional source(s) of the jet mechanical power. For the cases of black hole accretion, a promising energy source would be the black hole spin, i.e., the Blandford-Znejak (BZ) mechanism \citep{Blandford1977}. There have been evidences supporting this jet power origin for Galactic black hole binaries (e.g., \citealt{Narayan2012}; but also see, e.g., \citealt{Russell2013}). From our analyses for NGC 4631 X4, the presumable jet requires additional energy source besides the accretion to provide sufficient mechanical power to inflate the bubble structure. Numerical simulations on super-Eddington accretions by \citet{Narayan2017} demonstrate that the total energy conversion efficiency (including both radiative and mechanical power) of ULXs can be as high as $\sim0.7$ when introducing the high black hole spin ($a_{*}=0.9$) and the ``magnetic arrested disk" (MAD; e.g.,  \citealt{Bisnovatyi1976,Narayan2003}) models, where the majority of energy is carried out in the form of mechanical power. The black hole spin energy is extracted into the jets via the BZ mechanism.

\section{Conclusions}\label{sec:conclusion}

We present an optical imaging study of the five brightest X-ray sources in NGC 4631, among which \citet{Soria2009} identified four ULXs (X1, X2, X4, X5). \textit{Chandra}/ACIS data are utilized to obtain precise astrometry and to identify possible optical counterparts from the \textit{HST}/ACS images. A broad-band and narrow-band imaging campaign with CFHT/MegaCam is carried out to search for the bubble structures around the X-ray sources and to investigate their accretion states.

The supersoft X1 has a large optical-to-X-ray positional error ($\approx0\farcs5$) due to its low counts during the \textit{Chandra} observation. The candidate optical counterparts and the surrounding stars of X1 span a wide range of ages from 5 Myr to 80 Myr in the CMD, suggesting that they are likely not physically associated. X3 resides in a stellar environment with the age range of $\sim50$--80 Myr, while its three candidate counterparts show similar ages. X4 has six optical counterpart candidates, all of which show the age range consistent with that of the surrounding stars at $\sim20$--80 Myr. X5 appears to be associated with a star group with the age of $\sim5$--10~Myr, which is typical for the star clusters related to ULXs \citep{Poutanen2013}. This young star group is a manifestation of the strong star forming activity in the starburst galaxy NGC 4631. We do not provide the CMD for X2 due to its high extinction. 

A bubble nebula with a size of $\sim130$~pc $\times$ 100~pc around X4 is firstly detected in our CFHT/MegaCam H$\alpha$ narrow-band image. Unlike many other ULXs residing in the interior of their respective bubbles, this ULX is located at the east edge. It appears the H$\alpha$ bubble originates from X4 and expands one-sided towards the west direction, reaching maximum luminosity in the outermost region. In contrast, the extended structure appears smaller in the \oiii\ image, while its brightest section is much closer to the ULX and located to the east. The \oiii/H$\alpha$ line ratio map suggests that the H$\alpha$ bubble is generated mainly by shock ionization, while the \oiii\ structure is illuminated by the ULX via photoionization. 

The X4 bubble has an average surface brightness of $19.64\pm0.01$ mag arcsec$^{-2}$ in the H$\alpha$ band. By matching the observed \oiii/H$\alpha$ line ratio, we estimate the bubble expansion velocity $v_s \sim 110$ km~s$^{-1}$ and the ambient ISM density $n\sim6$~cm$^{-3}$ using the MAPPINGS V code. With these parameters, we constrain the mechanical power to inflate the bubble being $\sim5\times10^{40}$ erg s$^{-1}$ and the bubble age of $\sim7\times 10^5$ yr. Furthermore, we demonstrate that for non- or mildly- relativistic wind alone to generate the observed bubble, the needed mass-loss rate would be too high to sustain the long-term accretion. Instead, in the case of a relativistic jet (with a bulk Lorentz factor $\Gamma\sim10$) to inflate the bubble, the mass-loss rate would decrease to a more realistic level of $\sim10^{-7}\ M_{\odot}\ \rm yr^{-1}$. Similar to the cases of a few microquasars found in the Milky Way and nearby galaxies (e.g., SS~433, NGC 7793 S26, and M83 MQ1), the estimated mechanical jet power of NGC 4631 X4 is above the Eddington limit for a stellar-mass black hole. The black hole spin is likely to contribute to the jet power via the BZ mechanism. 

For future perspectives, optical spectroscopy, especially those with the integral-field instruments, will provide the bubble expansion velocity field and flux ratio map for a variety of emission lines, from which a more precise estimate of the mechanical power can be obtained. High-resolution X-ray spectroscopy will enable the measurement of outflow velocity, while deeper radio imaging with high angular resolution could reveal the ULX jet. With all these combined, we can derive a more reliable mass-loss rate of the outflow and further constrain the accretion models of ULXs.  

%\section{Software and third party data repository citations} \label{sec:cite}

\begin{figure}
\epsscale{1.2}
\plotone{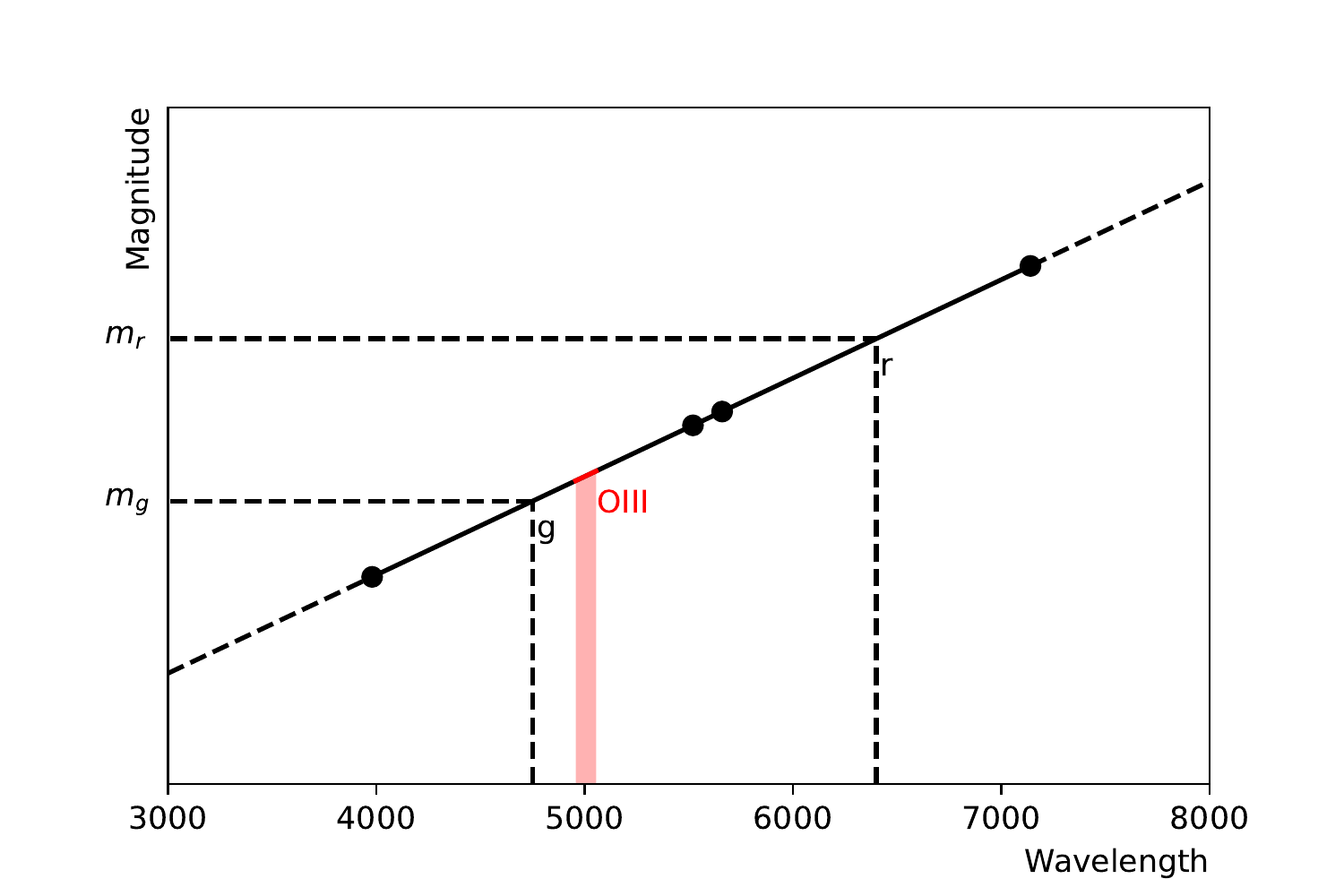}
\caption{The supposed linear relation between continuum magnitude and wavelength. The two pairs of black dots mark the boarders of the $g$ and $r$ bands of CFHT/MegaCam. The continuum in the \oiii\ band is illustrated by the red shaded area.}\label{fig:OiiiCtn}
\end{figure}

\begin{acknowledgments}
We thank S. Gwyn for processing the CFHT/MegaCam data with \verb+MegaPipe+ and S. Prunet for the help on imaging stacking. We thank A. Boselli and M. Fossati for helpful discussions on the continuum subtraction of H$\alpha$ images. We also thank S. Feng and Z. Li for archival VLA data enquiry. J.G. thank the CFHT staff for their hospitality during her visit to CFHT. This work is supported by the National Natural Science Foundation of China (grant No. U1938105, 12033004, U2038103) and the science research grants from the China Manned Space Project with NO. CMS-CSST-2021-A05 and CMS-CSST-2021-A06. 

This research uses data obtained through the Telescope Access Program (TAP), which has been funded by the TAP member institutes. Based on observations obtained with MegaPrime/MegaCam, a joint project of CFHT and CEA/DAPNIA, at the Canada-France-Hawaii Telescope (CFHT) which is operated by the National Research Council (NRC) of Canada, the Institut National des Sciences de l'Univers of the Centre National de la Recherche Scientifique of France, and the University of Hawaii. The observations at the Canada-France-Hawaii Telescope were performed with care and respect from the summit of Maunakea which is a significant cultural and historic site. 
Based on observations made with the NASA/ESA Hubble Space Telescope, and obtained from the Hubble Legacy Archive, which is a collaboration between the Space Telescope Science Institute (STScI/NASA), the Space Telescope European Coordinating Facility (ST-ECF/ESAC/ESA) and the Canadian Astronomy Data Centre (CADC/NRC/CSA). The data described here may be obtained from the MAST archive at \dataset[doi:10.17909/T9RP4V]{https://dx.doi.org/10.17909/T9RP4V}.
This research has made use of data obtained from the Chandra Data Archive and the Chandra Source Catalog, and software provided by the Chandra X-ray Center (CXC) in the application packages CIAO and Sherpa.
\end{acknowledgments}

\facilities{CFHT/MegaCam, \textit{HST}/ACS, \textit{Chandra}/ACIS}

\software{Astropy \citep{astropy13,astropy18}, 
          CIAO \citep{Fruscione2006},
          Dolphot \citep{Dolphin2000},
          Matplotlib \citep{Hunter2007},
          NumPy \citep{Harris2020},
          Pandas \citep{mckinney-proc-scipy-2010},
          PyRAF \citep{PyRAF2012},
          SCAMP \citep{Bertin2006},
          SciPy \citep{Virtanen2020},
          SExtractor \citep{Bertin1996},
          SWarp \citep{Bertin2010}.
          }

\appendix
%\begin{appendices}
\section{Subtracting the continuum from the \oiii\ band images} \label{Oiii continuum}

To remove the $g$-band contribution from the \oiii\ images, we assume the magnitude of continuum at a given wavelength is linearly correlated to this wavelength $\lambda$ in the range of the $g$ and $r$ bands (Figure~\ref{fig:OiiiCtn}), i.e., the continuum follows a power-law spectral model ($f \propto \lambda^{-\alpha}$). Then the $g$-band part in \oiii\ can be described in the equation,
\begin{equation}
    \frac{m_r-m_g}{\lambda_r-\lambda_g} = \frac{m_{g/\rm O\sc III}-m_g}{\lambda_{\rm O\sc III}-\lambda_g} .
\end{equation}
With replacing the corresponding central wavelength in the filters of Megacam ($\lambda_g$ = 4750\  \AA, $\lambda_{\rm O\sc III}$ = 5006\  \AA, $\lambda_r$ = 6400\ \AA), the equation can be transformed to, 
\begin{equation}
     m_{g/\rm O\sc III}\ \approx\ m_g - 0.155\times (m_g-m_r).
\end{equation}
It is worth noting that this relation is only valid to the images generated by \verb+MegaPipe+ for which the counts have been normalized for each band. The continuum component for each pixel can be derived after the equation is applied pixel by pixel. 

%\end{appendices}

\bibliographystyle{aasjournal}


\end{CJK*}
\end{document}